\documentclass[11pt,a4paper]{article}
\pdfoutput=1
\usepackage{jheppub}
\usepackage{slashed}
\usepackage{float}
\usepackage{bm}
\mathchardef\mhyphen="2D

\title{Drell-Yan process with jet vetoes: breaking of generalized factorization}
\author{Mao Zeng}
\affiliation{C.N.\ Yang Institute for Theoretical Physics, Stony Brook University,\\ Stony Brook, New York 11794--3840, USA}
\emailAdd{mao.zeng@stonybrook.edu}
\preprint{YITP-SB-15-28}
\abstract{Resummation of hadron collision cross sections, when the measurement imposes a hierarchy of scales, relies on factorization. Cancellation of Glauber / Coulomb gluons is a necessary condition for factorization. For Drell-Yan-like processes, the known proofs of cancellation of Glauber gluons are not applicable when jet vetoes are introduced, via jet algorithms or event shape variables such as the beam thrust. A priori, this does not rule out the possibility that an unknown new cancellation mechanism exists, or the possibility that a generalized factorization formalism is correct. To resolve the questions, we construct a direct counter-example in QCD with scalar quarks, contradicting any form of factorization in which the two collinear sectors are decoupled from each other. In the counter-example, decoupling of the two collinear sectors implies zero dependence of the beam thrust distribution on the longitudinal spin of the incoming hadrons, but we find a non-zero spin asymmetry at leading power due to Glauber gluons exchanged between spectators. We discuss implications for resumming large logarithms from jet vetoes.}


\arxivnumber{1507.01652}

\begin{document}
\maketitle
\flushbottom

\section{Introduction}
For the Drell-Yan (DY) process, the cancellation of Glauber gluons was a major difficulty in proving factorization \cite{Bodwin:1984hc, Collins:1985ue, Collins:1988ig}. In QCD factorization at leading power, by the Collins-Soper-Sterman (CSS) formalism \cite{Collins:1989gx, Sterman:1995fz, Collins:2011zzd} or soft collinear effective theory (SCET) \cite{Bauer:2000ew, Bauer:2000yr, Bauer:2001ct, Bauer:2001yt}, soft gluons decouple from the dynamics of collinear particles through the eikonal approximation. However, the eikonal approximation is not applicable to soft gluons whose momenta are dominated by transverse components, called Glauber / Coulomb gluons. This issue is of direct relevance for the resummation of jet veto logarithms at hadron colliders \cite{Banfi:2012yh, Banfi:2012jm, Tackmann:2012bt, Stewart:2013faa, Becher:2012qa, Becher:2013xia}, which requires re-factorization of the Drell-Yan cross section when the jet veto scale is much lower than the hard scale.

In the simpler cases of the Sudakov form factor and semi-inclusive deeply inelastic scattering (SIDIS), cancellation of Glauber gluons is achieved at the amplitude level, by deforming integration contours \cite{Collins:1981ta} away from the Glauber region. In such cases, if the Glauber region is included in the calculation, the cancellation happens between the Glauber region and the subtraction of overlap between the Glauber region and other regions \cite{Bauer:2010cc,Collins:2011zzd,Donoghue:2014mpa}. Such cancellation is by no means automatic; it places strong constraints \cite{Collins:2011zzd, Collins:2004nx} on the choice of Wilson lines in the definition of the soft and collinear functions, including the directions (past-pointing / incoming versus future-pointing / outgoing) and possible rapidity regularization.\footnote{The compatibility between contour deformation and rapidity regularization by off-lightcone Wilson lines \cite{Collins:2011zzd} has been studied in the aforementioned references, but it should be possible to extend the studies to other rapidity regulators \cite{Becher:2011dz,Chiu:2012ir} used in the SCET literature, which bear more resemblance to dimensional regularization.}

For the Drell-Yan process, the cancellation is more involved, due to the presence of both initial-state and final-state poles, in both collinear sectors. Incoming Wilson lines are chosen, to be compatible with contour deformation away from initial-state poles (which is responsible for the sign flip of the Sivers function between DY and SIDIS \cite{Collins:1992kk}), while final-state poles that obstruct contour deformation are canceled after summing over cuts and integrating over certain momentum components. While there are earlier proofs of cancellation of Glauber gluons, the CSS proof in \cite{Collins:1988ig}, based on both plus- and minus- lightcone ordered perturbation theory (LCOPT), is the most powerful one, because its applicability extends beyond leading-twist massless parton scattering \cite{Qiu:1990xy}, with important phenomenological applications to, e.g.\@ subleading-twist quarkonium production \cite{Kang:2014tta}.

The CSS proof originally required integrating over the transverse momenta of the partons, but it was subsequently realized that in the Feynman gauge, the proof carries through \cite{Laenen:2000ij, Collins:2011zzd} for transverse momentum dependent (TMD) factorization for the Drell-Yan process. However, for factorization of ``isolated" Drell-Yan production with measured hadronic event shape variables such as transverse energy \cite{Papaefstathiou:2010bw, Tackmann:2012bt, Grazzini:2014uha} and beam thrust \cite{Stewart:2009yx, Stewart:2010pd, Berger:2010xi} (see also \cite{Kang:2015moa}), the existing proofs are not applicable, as shown in \cite{Gaunt:2014ska} which explored connections with multi-parton interactions.

In fact, one of the crucial last steps of the CSS proof is integrating over the virtuality of the active partons (after summing over cuts in LCOPT).
This step is directly broken by a measurement of the beam thrust variable, as the factorization proposed in \cite{Stewart:2009yx} involves the virtuality-dependent PDF, also called the beam function. To address the miscancellation of Glauber gluons, Ref.\ \cite{Gangal:2014qda} introduced new jet veto observables that are designed to be less sensitive to such factorization-violating effects.

We will borrow the terminology ``generalized factorization" for hadron-hadron collisions proposed in a slightly different context, TMD factorization. Generalized factorization is to be distinguished from ``standard factorization" for hadron-hadron collisions, the latter of which assumes cancellation of Glauber gluons and always defines soft and collinear functions using past-pointing / incoming Wilson lines carrying the color charges of the active partons. For example, the factorization of beam thrust in \cite{Stewart:2009yx} using SCET should be characterized as standard factorization.\footnote{The equivalence of leading-power soft collinear factorization derived from traditional QCD methods and from SCET, for sufficiently inclusive observables that guarantee the cancellation of Glauber gluons, has been demonstrated extensively, for example in \cite{Lee:2006nr, Idilbi:2007ff, Idilbi:2007yi, Sterman:2013nya, Almeida:2014uva}} Generalized factorization, in a narrow sense, involves modification of Wilson lines in collinear and soft functions \cite{Bomhof:2004aw, Bacchetta:2005rm, Bomhof:2006dp, Vogelsang:2007jk}, but in a general sense, can be any factorization with a sensible spin structure \cite{Rogers:2013zha}.\footnote{For example, the large component of the Dirac spinor of a collinear quark should be projected out, and gluons that enter the hard scattering should carry two possible transverse polarizations.} In the model field theory considered in this paper, the colored active partons and the produced heavy uncolored particles are scalars carrying no spin indices, and generalized factorization is liberally defined as any factorization of the cross section into a product / convolution of scalar hard, collinear, anti-collinear, and possibly soft functions. We will make the definition precise in Section \ref{sec:Def}.

To disprove a leading-power factorization statement, it is sufficient to demonstrate, at some fixed order in $\alpha_s$, that the leading-power part of the cross section contradicts the prediction from factorization. Since QCD factorization relies on arguments that are applicable to \emph{all} unbroken gauge theories, irrespective of e.g.\@ gauge groups and matter contents, it is sufficient to find a contradiction in a model field theory that allows easy calculation. Model field theories involving polarized scattering, again in the slightly different context of TMD factorization, have been used to show the violation of both standard factorization and generalized factorization for hadron production at small transverse momentum \cite{Collins:2007nk, Collins:2007jp, Rogers:2010dm, Rogers:2013zha}. In particular, Ref.\ \cite{Rogers:2013zha} exploited discrete symmetries to study the violation of factorization, and this approach will be adopted in our study.

In this paper, we will consider Drell-Yan-like scattering in a model field theory, and study spin asymmetries in the doubly-differential beam thrust distribution. For this special model and observable, the vast majority of diagrams vanish, allowing a clean calculation of factorization-violating effects. The goal of the study is two fold. First, we would like to give an explicit demonstration that standard factorization is violated, which was shown by \cite{Gaunt:2014ska} to be extremely likely. Second, we would like to show that it is not possible for a generalized factorization theorem to hold, since decoupling of the two collinear sectors necessarily leads to zero spin asymmetry, while the calculations in this paper find a non-zero spin asymmetry. The non-cancellation of Glauber gluons found in this paper only happens above the jet veto scale, so collinear factorization is still valid if the jet veto scale is perturbative, but our ability to resum large logarithms in the hard scattering function will be compromised.

Ref.\ \cite{Gangal:2014qda} introduced new jet algorithm-based observables, such as the ``jet beam thrust'', which are designed to be less sensitive to Glauber effects, while preserving the rapidity-dependent nature of the beam thrust variable. Other recent research \cite{Collins:2007nk, Collins:2007jp, Rogers:2010dm, Rogers:2013zha, Forshaw:2006fk, Forshaw:2008cq, Catani:2011st, Forshaw:2012bi, StewartTalk} investigated the violation of QCD factorization in contexts other than, or wider than, the Drell-Yan process. For perturbative resummation, the logarithmic order at which factorization-violating effects start has been discussed for top quark pair production at low transverse momentum \cite{Li:2013mia, Catani:2014qha} and dijet event shapes \cite{Banfi:2010xy}.

This paper is organized as follows. In Section \ref{sec:model} we give a description of the model field theory and the observable measured in our thought experiment. In Section \ref{sec:proof} we explain why a non-zero spin asymmetry would contradict both standard factorization and generalized factorization. In Section \ref{sec:vanish} we check that up to $\mathcal O(\alpha_s^2)$, factorizable diagrams, as well as the vast majority non-factorizable diagrams, do not contribute, due to cancellations associated with this special observable. In Section \ref{sec:nonzero} we evaluate the only $\mathcal O(\alpha_s^2)$ cut diagram left, which involves one Glauber gluon exchanged on each side of the cut. The evaluation gives a non-zero spin asymmetry, which is the desired result. Some discussions are given in Section \ref{sec:dis}.
\section{The model for showing factorization breaking}
\label{sec:model}
\subsection{Model field theory}
\begin{figure}[tb]
\begin{centering}
\includegraphics[scale=0.6]{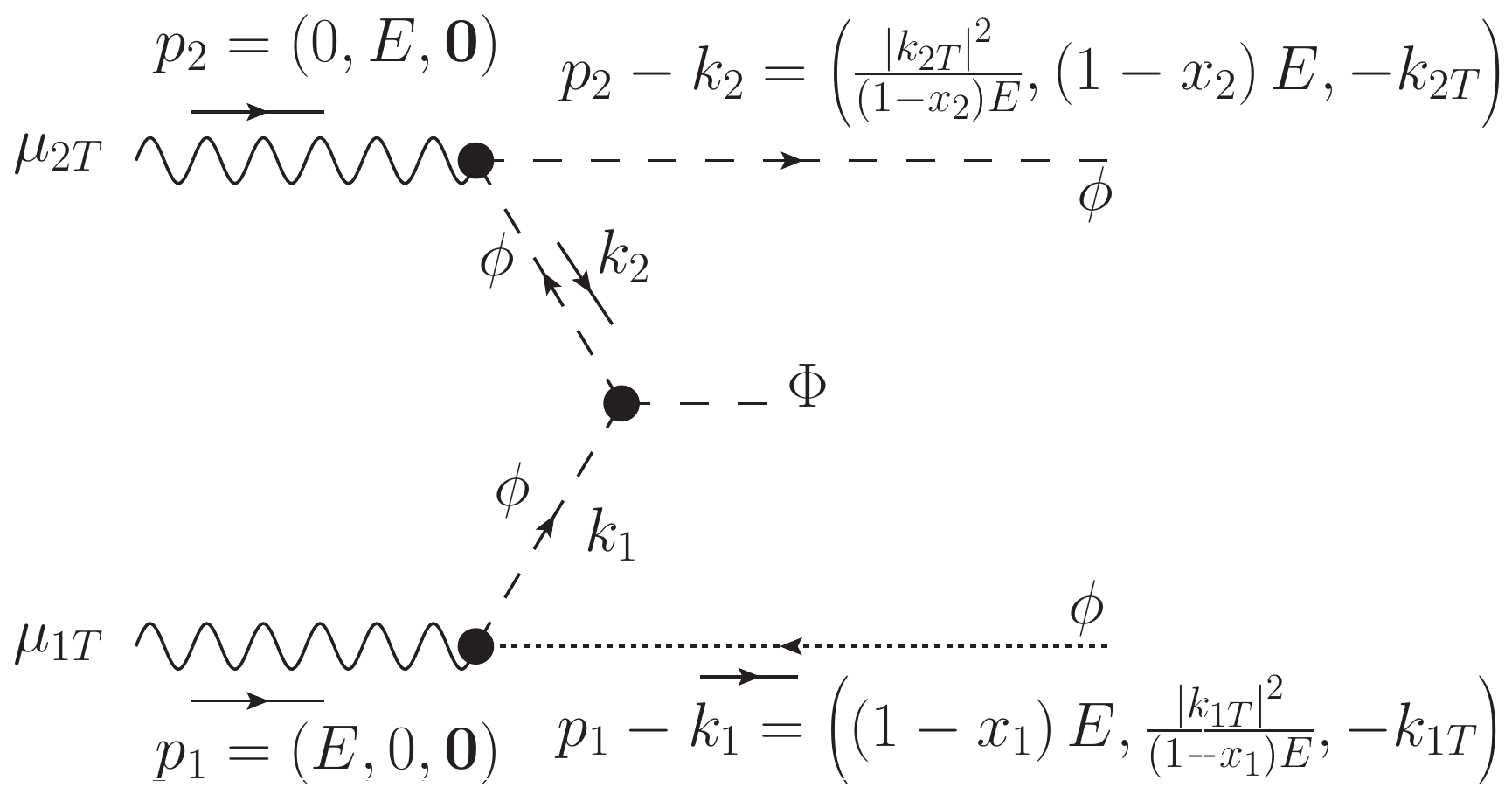}
\par\end{centering}
\caption{The leading-order Feynman diagram for $\gamma + \gamma \rightarrow \Phi + X$ in the model theory. Here $X = \phi + \phi^*$.} 
  \label{fig:DY-LO}
\end{figure}

We consider QCD with a hypothetical massless complex scalar quark $\phi$ (instead of a Dirac fermion quark as in real QCD) under the fundamental representation of $SU(3)$. The scalar quark also carries an electric charge of $+e$. The produced heavy particle $\Phi$ (analogous to $\gamma^*$ and $Z^0$ in the real Drell-Yan process) is a neutral color-singlet scalar with mass $M$, and couples to the light scalar quarks via the interaction Lagrangian $-g_{\Phi} \Phi \phi^* \phi$. The photon $\gamma$ serves the role as a "hadron" which can split into an active scalar quark that participates in hard scattering, and a spectator scalar quark going down the beam pipe. We consider the Drell-Yan process $\gamma + \gamma \rightarrow \Phi + X$. The lowest-order diagram is $\gamma + \gamma \rightarrow \Phi + \phi + \phi^*$, shown in Fig.\@ \ref{fig:DY-LO}.
\subsection{Observable - spin asymmetry in doubly differential beam thrust}
We will use the beam thrust variable \cite{Sterman:2005bf, Stewart:2009yx}
\begin{equation}
\tau_B \equiv \frac 1 M \sum_{i} |p_{T_i}| e^{-|y_i-y|},
\label{eq:tauB}
\end{equation}
where $y$ and $M$ are the rapidity and invariant mass of the Drell-Yan pair (actually a heavy scalar in our model), and the index $i$ runs over every detected hadron. Requiring $\tau_B \ll 1$ strongly restricts hadronic activity, especially in the central rapidity region. It is useful to consider the \emph{doubly-differential distribution} \cite{Stewart:2009yx} in $(\tau_{R}, \tau_{L})$, where $\tau_{R}$ receives contribution from only the right hemisphere ($y_i > y$) and $\tau_{L}$ receives contribution from only the left hemisphere ($y_i < y$).

Even though the arguments in this paper also apply to jet algorithm-based vetoes, for concreteness we will consider the doubly-differential beam thrust distribution from the scattering of two photons. Our thought experiment involves incoming photons with several polarization configurations, from which we obtain the double spin asymmetry in the distribution.
We will do the calculation for a heavy scalar $\Phi$ produced at rapidity $y=0$, and with the left and right hemispheres having the same beam thrust $\tau_{R} = \tau_{L} = \frac 1 2 \tau_B \ll 1$. The phase space integral for the square of the amplitude in Fig. \ref{fig:DY-LO} is, in the approximation that $k_1$ is plus-collinear and $k_2$ is minus-collinear,
\begin{align}
&\quad \frac {d^3 \sigma} {d \tau_{R} d \tau_{L} d y} \bigg|_{y=0, \, \tau_{R}=\tau_{L}=\tau_B/2} \nonumber \\
&=\frac{1}{2E^2} \int \frac{d x_1}{2(2\pi) \left( 1- x_1 \right) } \int \frac{d x_2}{2(2\pi) \left( 1- x_2 \right) } \int \frac {d^2 k_{1T}} { (2\pi)^2 } \int \frac {d^2 k_{2T}} { (2\pi)^2 } \left| \mathcal M \right|^2 \nonumber \\
&\quad 2\pi \delta \left( x_1 x_2 E^2 - M^2 \right) \delta \left( \frac 1 2 \ln \frac {x_1}{x_2} \right) \delta \left( \frac {\tau_B}{2} - \frac { \left| k_{1T} \right|^2 } {M \left( 1-x_1 \right) E } \right) \delta \left( \frac {\tau_B}{2}  - \frac { \left| k_{2T} \right|^2 } { M \left( 1-x_2 \right) E } \right)
\label{eq:phaseSpace}
\end{align}
The delta functions in Eq.\ \eqref{eq:phaseSpace} force
\begin{align}
x_1 = x_2 &= \frac {M}{E} \equiv x, \label{eq:x1x2x} \\
\left| k_{1T} \right| = \left| k_{2T} \right| &= \sqrt{ \frac {\tau_B (1-x) } {2} } \equiv \left| k_{0T} \right|,
\label{eq:k1k2k0}
\end{align}
and Eq. \eqref{eq:phaseSpace} simplifies to
\begin{align}
&\quad \frac {d^3 \sigma} {d \tau_{R} d \tau_{L} d y} \bigg|_{y=0, \, \tau_{R}=\tau_{L}=\tau_B/2} \nonumber \\
&= \frac {M^2}{16\pi E^2} \int \frac {d^2 k_{1T}} { (2\pi)^2 } \int \frac {d^2 k_{2T}} { (2\pi)^2 } \delta \left( \left| k_{1T} \right|^2 - \left| k_{0T} \right|^2 \right) \delta \left( \left| k_{2T} \right|^2 - \left| k_{0T} \right|^2 \right) \left| \mathcal M \right|^2.
\label{eq:phaseSpace1}
\end{align}

For the scattering of two photons, we define the \emph{absolute} double spin asymmetry as
\begin{equation}
\sigma_{\rm asym} = \left( \sigma_{\uparrow \downarrow} + \sigma_{\downarrow \uparrow} - \sigma_{\uparrow \uparrow} - \sigma_{\downarrow \downarrow} \right) / 4,
\end{equation}
where up and down arrows denote right and left polarizations. The \emph{relative} double spin asymmetry is defined as the above expression divided by the unpolarized cross section
\begin{equation}
\sigma_{\rm unpol} = \left( \sigma_{\uparrow \downarrow} + \sigma_{\downarrow \uparrow} + \sigma_{\uparrow \uparrow} + \sigma_{\downarrow \downarrow} \right) / 4.
\end{equation}
When a photon traveling in the $z$ direction is right polarized the polarization vector is $\epsilon_{\uparrow}=\left( \epsilon^0_{\uparrow}, \epsilon^x_{\uparrow}, \epsilon^y_{\uparrow}, \epsilon^z_{\uparrow} \right) = (0,1,i,0) / \sqrt 2$. So $\epsilon^\mu_{\uparrow} \epsilon^{*\nu}_{\uparrow} = \left( -g^{\mu_T \nu_T} - i \epsilon^{\mu_T \nu_T} \right) / 2$, where $\epsilon^{xy}=-\epsilon^{yx}=1$. Similarly, for a left-polarized photon, $\epsilon^\mu_{\downarrow} \epsilon^{*\nu}_{\downarrow} = \left( -g^{\mu_T \nu_T} + i \epsilon^{\mu_T \nu_T} \right) / 2$. The half difference between the polarization sums is
\begin{equation}
\frac 1 2 \left( \epsilon^\mu_{\downarrow} \epsilon^{*\nu}_{\downarrow} - \epsilon^\mu_{\uparrow} \epsilon^{*\nu}_{\uparrow} \right) = \frac 1 2 i \epsilon^{\mu_T \nu_T},
\label{eq:spin-factor}
\end{equation}
To obtain the (absolute) double spin asymmetry in the distribution, we can replace $ | M |^2$ in Eq.\@ \eqref{eq:phaseSpace1} by
\begin{equation}
|\mathcal M|^2_{\rm asym} = \frac 1 4 \epsilon_{\mu_{1T} \, \nu_{1T}} \epsilon_{\mu_{2T} \, \nu_{2T}} \mathcal M^{\mu_{1T} \, \mu_{2T}} \mathcal M^{\nu_{1T} \, \nu_{2T}},
\label{eq:absoluteAsym}
\end{equation}
while to obtain the unpolarized cross section, we use the averaged squared matrix element
\begin{equation}
|\mathcal M|^2_{\rm unpol} = \frac 1 4 g_{\mu_{1T} \, \nu_{1T}} g_{\mu_{2T} \, \nu_{2T}} \mathcal M^{\mu_{1T} \, \mu_{2T}} \mathcal M^{\nu_{1T} \, \nu_{2T}},
\label{eq:relativeAsym}
\end{equation}
\section{Outline of the proof by contradiction}
\label{sec:proof}
\subsection{Definition of standard factorization}
\label{sec:Def}
By standard factorization, we mean factorization derived by assuming the cancellation of spectator-spectator Glauber gluon exchanges. For the doubly differential beam thrust distribution (in the left and right hemispheres) in 
Drell-Yan production of a system of invariant mass $M$, at hadronic c.o.m.\@ collision energy $E$, the factorization formula in double Laplace moment space is \cite{Stewart:2009yx, Berger:2010xi}
\begin{align}
&\quad \int_0^\infty d \tau_{R}\, e^{ -N_R \, \tau_{R} } \int_0^\infty d \tau_{L}\, e^{ -N_L \, \tau_{L} } \frac {d^3 \sigma \left( H_1 + H_2 \rightarrow \Phi + X \right) } {d \tau_{R} d \tau_{L} d y} \bigg|_{y=0} \nonumber \\
&= H \left( \frac{M^2} {\mu_f^2}, \alpha_s \left( \mu_f \right) \right) \tilde B_1 \left( \frac {M^2} {N_R \, \mu_f^2}, x, \alpha_s \left( \mu_f \right) \right) \tilde B_2 \left( \frac {M^2} {N_L \, \mu_f^2}, x, \alpha_s \left( \mu_f \right) \right) \nonumber \\
&\times \tilde S \left( \frac{M^2} {N_R^2 \, \mu_f^2}, \frac{M^2} {N_L^2 \, \mu_f^2}, \alpha_s \left( \mu_f \right) \right) + \mathcal O \left( \frac 1 {N_R}, \frac 1 {N_L} \right),
\label{eq:factorization}
\end{align}
where $x= M / E$ is the Bjorken variable for both collinear sectors, as we imposed $y=0$. $H$ is the hard function. $\tilde B_1$ and $\tilde B_2$ are the two moment-space collinear functions, also called beam functions in the literature, for the collinear sectors initiated by the incoming hadrons $H_1$ and $H_2$, respectively. $\tilde S$ is the moment-space soft function. The beam function is the ``virtuality-dependent PDF" \cite{Stewart:2009yx}, originally defined using SCET fields with implicit zero-bin subtraction \cite{Manohar:2006nz}. In Appendix \ref{sec:beam} we define the beam function using the scalar QCD fields used in this paper.

\subsection{Definition of generalized factorization}
We first re-write Eq.\ \eqref{eq:factorization}, dropping explicit dependence on $\mu_f$ and $\alpha_s (\mu_f)$, and combining $H$ and $\tilde S$ into a function $\tilde R$,
\begin{align}
&\quad \int_0^\infty d \tau_{R} e^{ -N_R \, \tau_{R} } \int_0^\infty d \tau_{L} e^{ -N_L \, \tau_{L} } \frac {d^3 \sigma \left( H_1 + H_2 \rightarrow \Phi + X \right) } {d \tau_{R} d \tau_{L} d y} \bigg|_{y=0} \nonumber \\
&= H \left( M^2 \right) \tilde B_1 \left( M^2, N_R, x\right) \tilde B_2 \left( M^2, N_L, x\right)  \tilde S \left( M^2, N_R, N_L \right) + \mathcal O \left( \frac 1 {N_R}, \frac 1 {N_L} \right) \nonumber \\
&= \tilde B_1 \left( M^2, N_R, x\right) \tilde B_2 \left( M^2, N_L, x\right)  \tilde R \left( M^2, N_R, N_L \right) + \mathcal O \left( \frac 1 {N_R}, \frac 1 {N_L} \right).
\label{eq:factorization1}
\end{align}

Generalized factorization is defined by Eq.\ \eqref{eq:factorization1}, with the following requirements. $\tilde B_1$, $\tilde B_2$, and $\tilde R$ are allowed to be arbitrary functions that may differ from how they are defined in standard factorization.\footnote{For example, $\tilde B_1$ in Eq.\ \eqref{eq:factorization1} can depend on $M^2$ and $N_R$ independently, while $\tilde B_1$ in Eq.\ \eqref{eq:factorization} is written in a form that can only depend on these two variables through the combination $M^2 / N_R$.}
The only dependence on the incoming state $H_1$, including the species of the hadron and the polarization, should be contained in $\tilde B_1$, and the same condition is imposed on $H_2$ and $\tilde B_2$. In other words, we require the two collinear sectors to be decoupled. The various functions on the R.H.S.\ of Eq.\ \eqref{eq:factorization1} generally still depend on additional auxiliary variables that are not shown, including the factorization scale $\mu_f$ and the directions of Wilson lines involved in the definition of these functions.

It is the aim of the next subsection to show that generalized factorization is incompatible with the result of this paper; standard factorization, a special case of generalized factorization, is thus also violated. Before doing so, we give an example of generalized factorization that differs from standard factorization. In the standard factorization formula, Eq.\ \eqref{eq:factorization}, the soft function is defined using incoming Wilson lines \cite{Stewart:2009yx}. As shown in \cite{Kang:2015moa}, up to $\mathcal O(\alpha_s^2)$, if outgoing Wilson lines are used instead, the soft function remains the same up to $\mathcal O(\alpha_s^2)$. By flipping the directions of Wilson lines in the soft function, we turn Eq.\ \eqref{eq:factorization} into a generalized factorization formula which coincides with standard factorization at the first few $\alpha_s$ orders, but differs at higher $\alpha_s$ orders.

\subsection{Violation of generalized factorization}
Under the generalized factorization formula Eq.\ \eqref{eq:factorization1}, the double longitudinal spin asymmetry in the factorized beam thrust distribution is, in schematic moment-space factorized form,
\begin{equation}
-\frac 1 4 \left( \tilde B_1^\uparrow - \tilde B_1^\downarrow \right) \left( \tilde B_2^\uparrow - \tilde B_2^\downarrow \right) \tilde R,
\label{eq:doubleAsymSchematic}
\end{equation}
while the corresponding expression for single longitudinal spin asymmetry is
\begin{equation}
\frac 1 2 H \left( \tilde B_1^\uparrow - \tilde B_1^\downarrow \right) \tilde B_2^{\rm unpol} \tilde R.
\label{eq:singleAsymSchematic}
\end{equation}
In Eqs.\@ \eqref{eq:doubleAsymSchematic} and \eqref{eq:singleAsymSchematic}, we used ``$\uparrow$", ``$\downarrow$", and ``unpol" to denote right polarization, left polarization, and no polarization, respectively.
It is an immediate consequence of parity conservation of our model theory that the single spin asymmetry in Eq.\ \eqref{eq:singleAsymSchematic} must be vanishing. Since the unpolarized scattering cross section is non-zero, the unpolarized beam function $\tilde B_2^{\rm unpol}$ in Eq.\@ \eqref{eq:singleAsymSchematic} cannot be vanishing (except at isolated points, assuming the function is analytic). So $\tilde B_1^\uparrow - \tilde B_1^\downarrow$ must be almost everywhere zero, which means everywhere zero if the function is analytic.\footnote{Indeed, for the special case of standard factorization, we can check from the definition of the beam function that $\tilde B_1$ and $\tilde B_2$ do not depend on the polarizations of the hadrons.} This implies that the double spin asymmetry, given in the schematic factorized form Eq.\@ \eqref{eq:doubleAsymSchematic}, must also vanish at leading power to all orders in $\alpha_s$. But we need to consider non-factorizable contributions from the Glauber region in spectator-spectator interaction, and its overlap with other regions (to be subtracted); if the sum of these contributions is non-zero, as we will find in the subsequent sections, we obtain a contradiction to generalized factorization.\footnote{We will actually show that the distribution is non-zero at some $ \left( \tau_B, \tau_L \right)$, but this is sufficient to imply that the distribution cannot be an identically vanishing function in moment space.}
\section{Vanishing diagrams}
\label{sec:vanish}
\subsection{Vanishing LO diagram}
\label{sec:one-Glauber}
We know that factorizable diagrams give a vanishing contribution to the double spin asymmetry in our model, but we will explicitly verify that the LO contribution from squaring the diagram in Fig.\@ \ref{fig:DY-LO} vanishes, to introduce notations and demonstrate cancellations that are also applicable to some non-factorizable diagrams.
For brevity, we write
\begin{align}
D \left( p \right) &\equiv \frac i {p^2 + i \epsilon}, \label{eq:D}\\
\epsilon \left( p_1 , p_2 \right) &\equiv \epsilon_{\mu_{1T}\, \mu_{2T}} \, p_1^{\mu_{1T}} p_2^{\mu_{2T}}.\label{eq:epsilon}
\end{align}
The diagram evaluates to, given that $p_1$ and $p_2$ have no transverse components,
\begin{align}
i \mathcal M_{\rm LO}^{ \mu_{1T} \, \mu_{2T} } &= (-i) g_\Phi 2ie\, k_1^{\mu_{1T}} (-2ie)\, k_2^{\mu_{2T}} D \left( k_1 \right) D \left(k_2 \right) \nonumber \\
&= -4i e^2 g_\Phi \, k_1^{\mu_{1T}} \, k_2^{\mu_{2T}} D \left( k_1 \right) D \left(k_2 \right).
\label{eq:MLO}
\end{align}
In this expression we omitted the color factor $\delta_{ab}$, with $a$ and $b$ being the color indices for the scalar / anti-scalar pair in the final state. Using Eq.\@ \eqref{eq:absoluteAsym}, the resulting double spin asymmetry in the squared matrix element is,
\begin{equation}
\frac 1 4 \epsilon_{\mu_{1T} \, \nu_{1T}} \epsilon_{\mu_{2T} \, \nu_{2T}} \mathcal M_{\rm LO}^{ \mu_{1T} \, \mu_{2T} } \left( \mathcal M_{\rm LO}^{ \nu_{1T} \, \nu_{2T} } \right)^* \propto \epsilon \left(k_1^T,k_1^T\right) \epsilon \left(k_2^T,k_2^T\right) = 0.
\end{equation}
In contrast, the unpolarized spin-summed squared matrix element is, using Eqs.\@ \eqref{eq:x1x2x} for zero rapidity,
\begin{equation}
| \mathcal M_{\rm LO}|^2_{\rm unpol} = 
2C_A \cdot \frac 1 4 g_{\mu_{1T} \, \nu_{1T}} g_{\mu_{2T} \, \nu_{2T}}  \mathcal M_{\rm LO}^{ \mu_{1T} \, \mu_{2T} } \left( \mathcal M_{\rm LO}^{ \nu_{1T} \, \nu_{2T} } \right)^* =2C_A \cdot 4e^4 g_\Phi^2 \frac{\left( 1-x \right)^4}{\left|k_{1T}\right|^2 \left|k_{2T}\right|^2},
\label{eq:M2-unpol}
\end{equation}
In this expression we include the color factor $C_A$ from the final-state color sum, and an overall factor of 2 to account for the possibility of reversing the complex scalar arrow (i.e.\@ swapping scalar and anti-scalar). The explicit momentum components in Fig.\@ \ref{fig:DY-LO} have been used to evaluate $D(k_1)$ and $D(k_2)$. Using Eq. \eqref{eq:M2-unpol} as the squared matrix element in the phase space integral Eq. \eqref{eq:phaseSpace1}, we obtain the LO unpolarized beam thrust distribution
\begin{equation}
\frac {d^3 \sigma_{\rm LO}} {d \tau_R d \tau_L dy} \bigg|_{y=0, \tau_R = \tau_L = \tau_B/2} = \frac {M^2} {16\pi E^2} 2C_A \cdot 4e^4 g\Phi^2 (1-x)^4 I_{\rm LO},
\label{eq:LOdistribution}
\end{equation}
where we defined
\begin{align}
I_{\rm LO} &= \int \frac {d^2 k_{1T}} { (2\pi)^2 } \int \frac {d^2 k_{2T}} { (2\pi)^2 }
\delta \left( \left| k_{1T} \right|^2 - \left| k_{0T} \right|^2 \right) \delta \left( \left| k_{2T} \right|^2 - \left| k_{0T} \right|^2 \right) \frac{1} {\left| k_{1T} \right|^2 \left| k_{2T} \right|^2} \nonumber \\
&=\frac 1 { (4\pi)^2 \left| k_{0T} \right|^4 }.
\label{eq:integralLO}
\end{align}
\subsection{Vanishing one-loop cut diagram}
\begin{figure}[tb]
\begin{centering}
\includegraphics[scale=0.6]{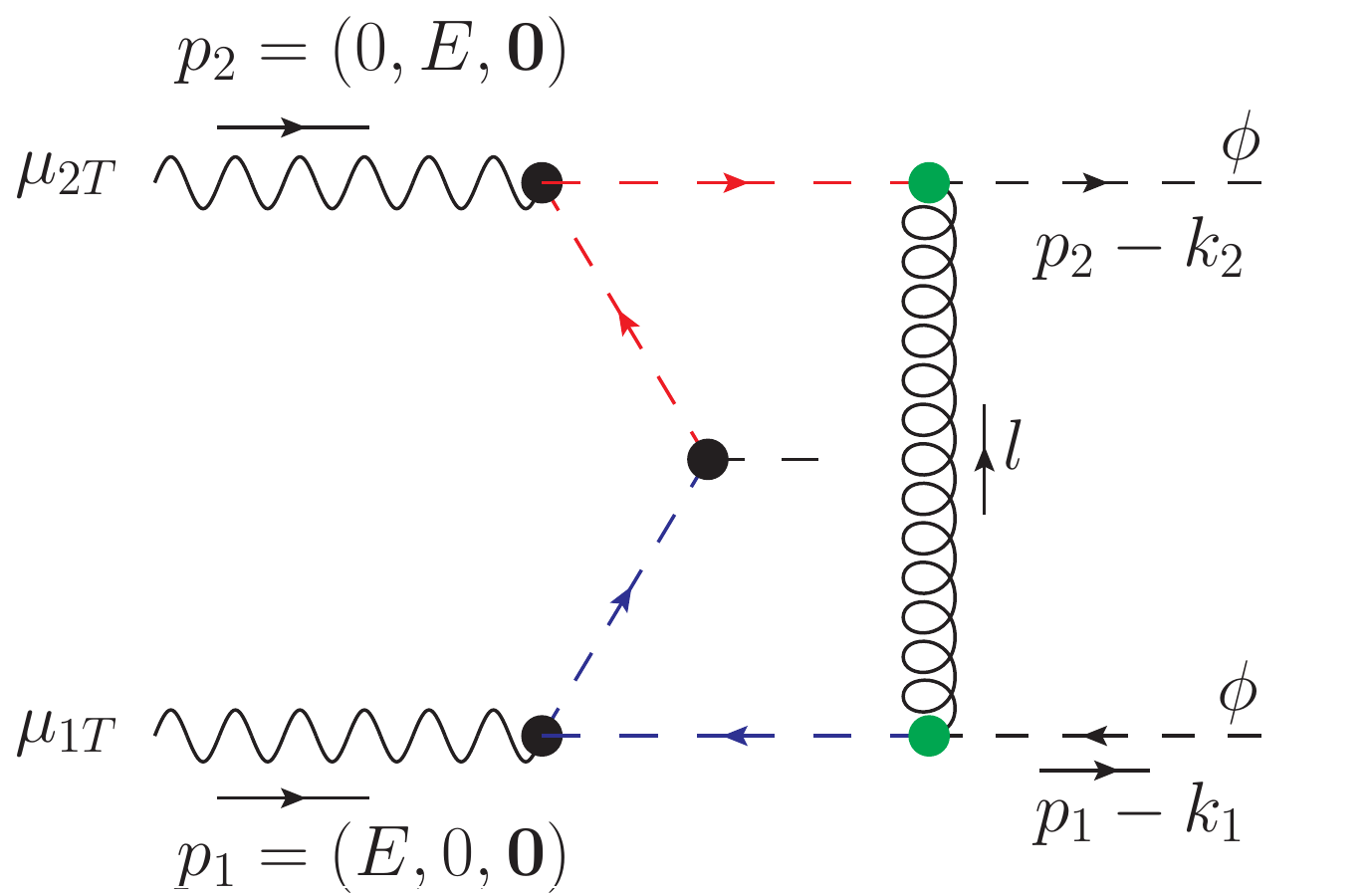}
\par\end{centering}
\caption{The one-Glauber exchange diagram in the model field theory.} 
  \label{fig:spectator}
\end{figure}

Consider the diagram Fig.\@ \ref{fig:spectator} in which the two spectator lines are connected by one gluon that is either a Glauber gluon or a normal soft gluon, \emph{in interference} with the complex conjugate of the LO diagram. We will show that the resulting contribution to the spin asymmetry has vanishing real and imaginary parts. For more general models and observables, the contribution is purely imaginary when the gluon has Glauber-like momentum, and cancels with the complex conjugate cut diagram \cite{Bodwin:1984hc, Gaunt:2014ska}.\footnote{A rare exception is single transverse spin asymmetry, for which factorization can be violated by the exchange of only one Glauber gluon, because the resulting imaginary contribution is multiplied by another imaginary factor from the Dirac trace with $\gamma^5$, to give a real contribution \cite{Collins:2007nk}.} But in our special model theory, even the imaginary contribution vanishes.
Fig.\@ \ref{fig:spectator} evaluates to, again noticing $p_1^T = p_2^T =0$ and omitting the color factor $C_F \delta_{ab}$,
\begin{align}
i \mathcal M_1^{ \mu_{1T} \, \mu_{2T} } &= -i g_\Phi \int \frac{d^4 l}{(2\pi)^4}\, 4e^2 g_s^2 \, \left( k_1^{\mu_{1T}} - l^{\mu_{1T}} \right) \, \left( k_2^{\mu_{2T}} + l^{\mu_{2T}} \right) \left( 2p_1 - 2k_1 + l \right) \cdot \left( 2p_2 - 2k_2 -l \right) \nonumber \\
&\quad D \left( l \right) D \left( p_1 - k_1 +l \right) D \left( p_2 - k_2 -l \right)
D \left( k_1 - l \right) D \left( k_2 + l \right) \\
& \approx -8i e^2 g_\Phi \, g_s^2 \left( p_1^+ - k_1^+ \right) \left( p_2^- -k_2^- \right)  \int d^4 l\, \left( k_1^{\mu_{1T}} - l^{\mu_{1T}} \right) \, \left( k_2^{\mu_{2T}} + l^{\mu_{2T}} \right) \nonumber \\
&\quad D \left( l \right) D \left( p_1 - k_1 +l \right) D \left( p_2 - k_2 -l \right)
D \left( k_1 - l \right) D \left( k_2 + l \right),
\label{eq:M1}
\end{align}
where we used approximation $l^+,l^- \ll Q$, applicable to both the Glauber and the normal soft region, after the ``$\approx$" sign.
Therefore, the interference between the diagram in Fig.\@ \ref{fig:spectator} and the complex conjugate of the LO diagram in Fig.\@ \ref{fig:DY-LO} is
\begin{align}
&\quad \epsilon_{\mu_{1T} \, \nu_{1T}} \epsilon_{\mu_{2T} \, \nu_{2T}} \mathcal M_1^{ \mu_{1T} \, \mu_{2T} } \left( \mathcal M_{\rm LO}^{ \nu_{1T} \, \nu_{2T} } \right)^* \nonumber \\
& \propto \int d^4 l \, \epsilon \left( k_1^T, l^T \right) \epsilon \left(k_2^T, l^T \right) D \left( l \right) D \left( p_1 - k_1 +l \right) D \left( p_2 - k_2 -l \right)
D \left( k_1 - l \right) D \left( k_2 + l \right),
\label{eq:LO-1}
\end{align}
where only $l_T$-dependent terms are shown after the proportional sign ``$\propto$".
Recall that we would like to measure the doubly-differential beam thrust distribution at some $\tau_{L} = \tau_{R}$, with the heavy particle $\Phi$ produced at zero rapidity. We still need to integrate the squared matrix element Eq.\@ \eqref{eq:LO-1} over the phase space of $k_1$ and $k_2$ with appropriate measurement functions. Consider a particular point in the $l$-integration volume in Eq.\@ \eqref{eq:LO-1}, for example, a point with $l_x \neq 0$, $l_y = 0$ without loss of generality. Then we can flip the sign of the $y$ component of $k_1$ without changing any terms in the integrand in Eq.\@ \eqref{eq:LO-1} except for flipping the sign of $\epsilon \left( k_1^T, l^T \right)$. Since jet veto observables are azimuthally symmetric and do not generate a preferred $y$-direction for $k_1$, Eq.\@ \eqref{eq:LO-1} gives a vanishing contribution to the doubly differential beam thrust distribution. At some general value of $l$, the needed change of variable is
\begin{equation}
k_1^T \rightarrow R_{l^T} \circ k_1^T = \left( k_1^x, k_1^y \right) - 2 \frac{ l^x k_1^y - l^y k_1^x } {| l^T |^2} \left( -l^y, l^x \right),
\label{eq:reflect}
\end{equation}
denoting a reflection of $k_1$ in the line through the origin in the $\pm l_T$ direction. The squared matrix element in Eq.\@ \eqref{eq:LO-1} has odd parity under this transformation, while the beam thrust variables $(\tau_{R}, \tau_{L})$ are invariant. Therefore the contribution vanishes after phase space integration.
\subsection{Vanishing two-loop cut diagrams}
\label{sec:2loopzero}
\begin{figure}[tb]
\begin{tabular}{cc}
  \includegraphics[width=65mm]{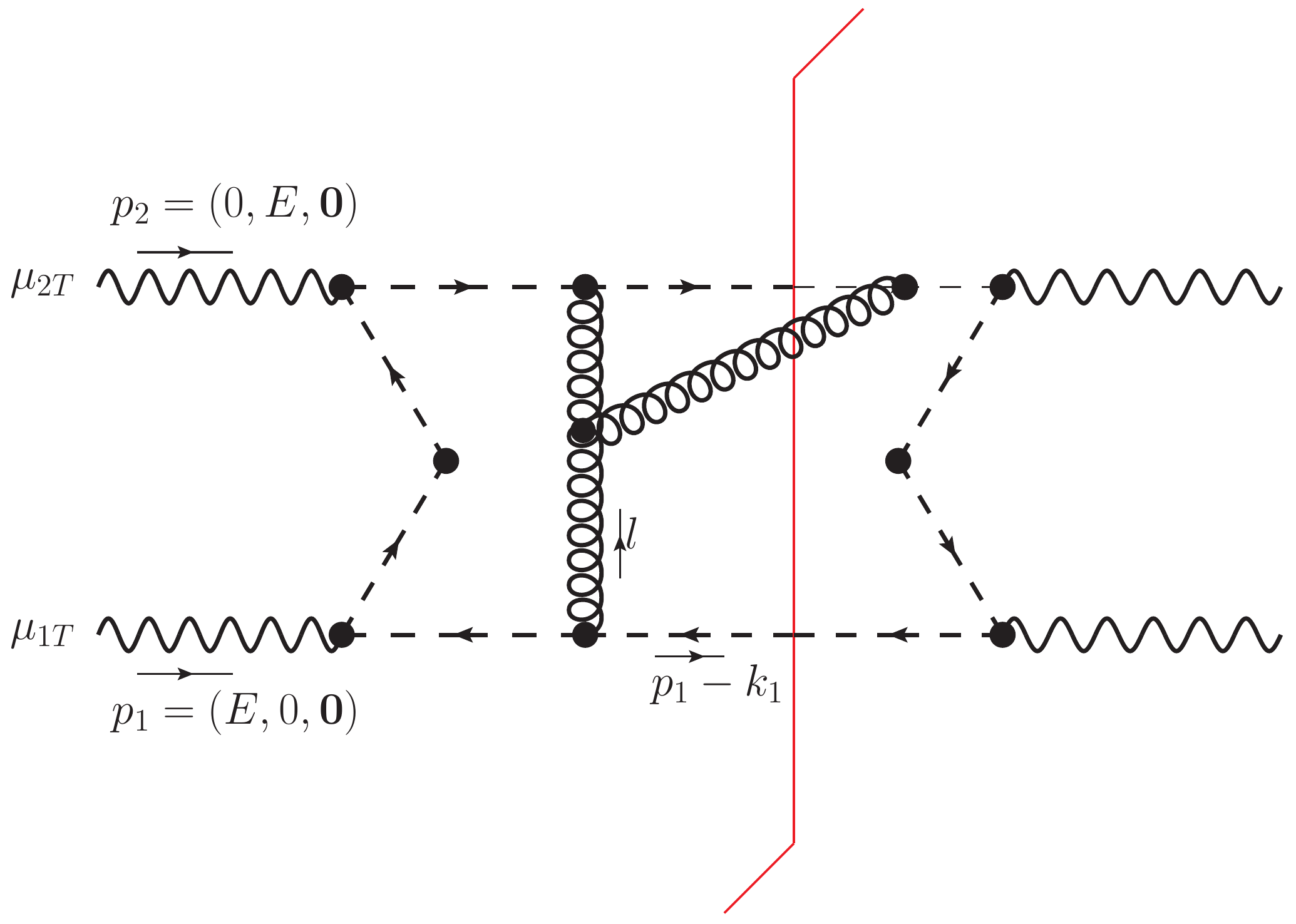} &   \includegraphics[width=65mm]{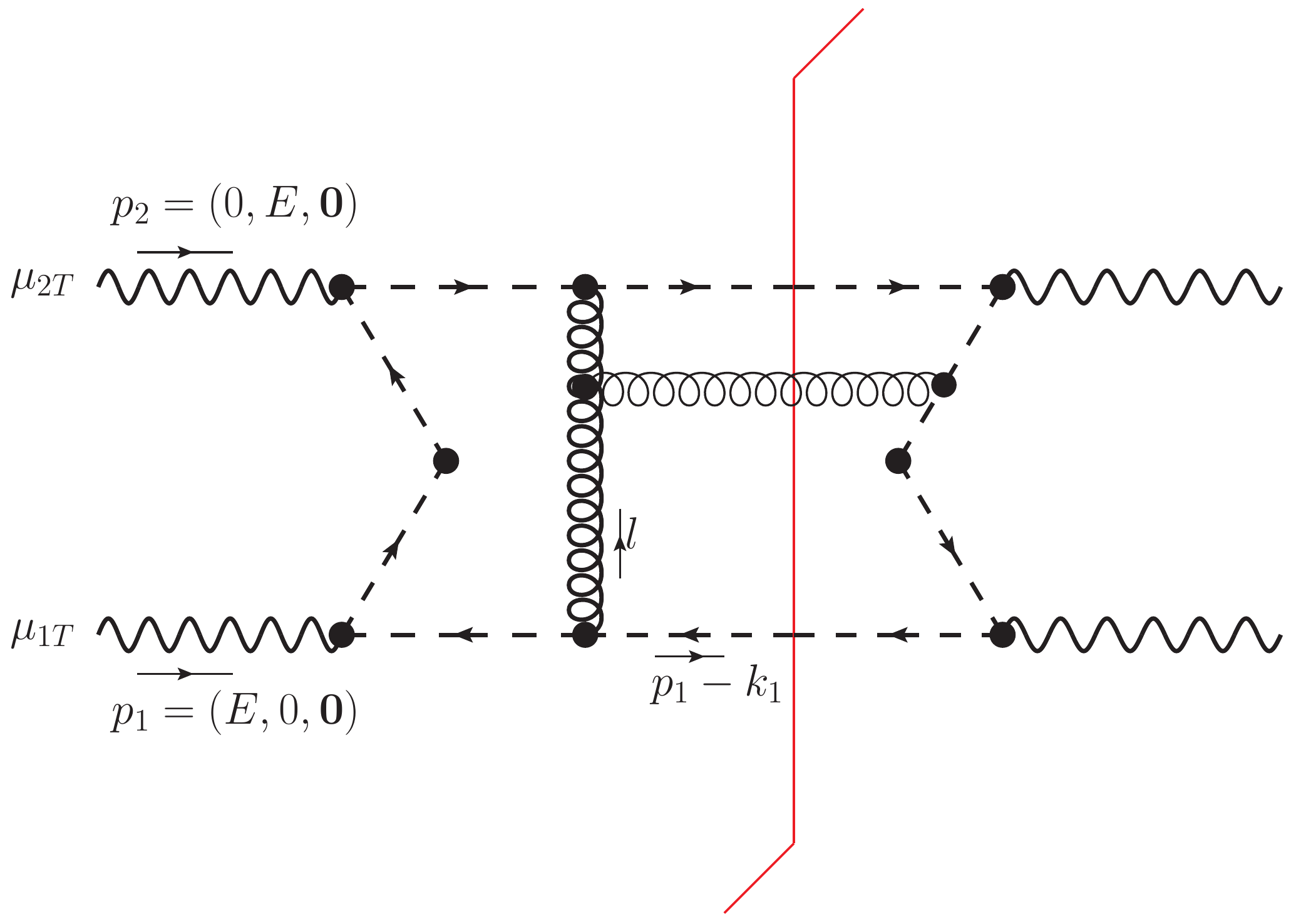} \\
(a)& (b) \\[6pt]
 \includegraphics[width=65mm]{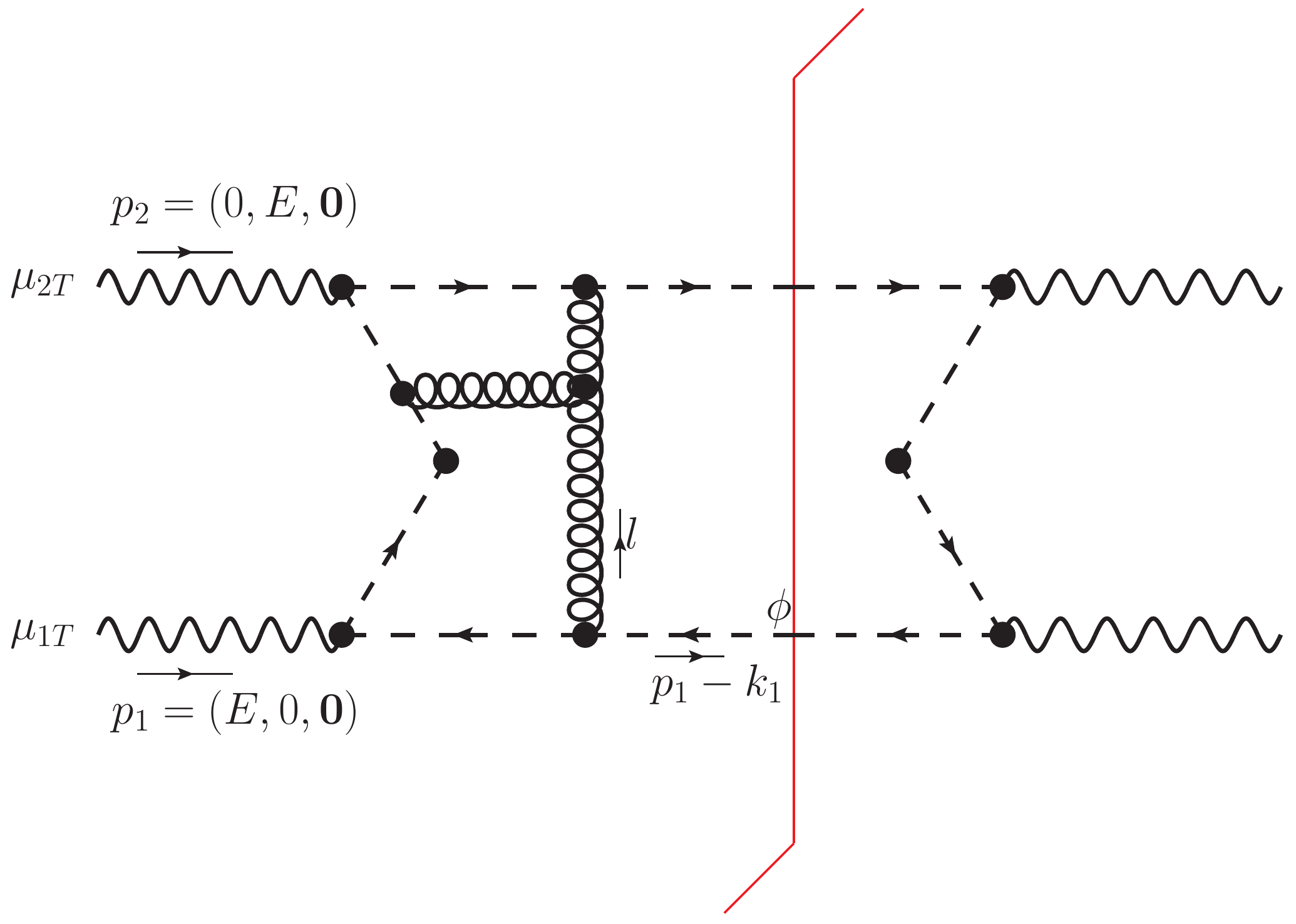} &   \includegraphics[width=65mm]{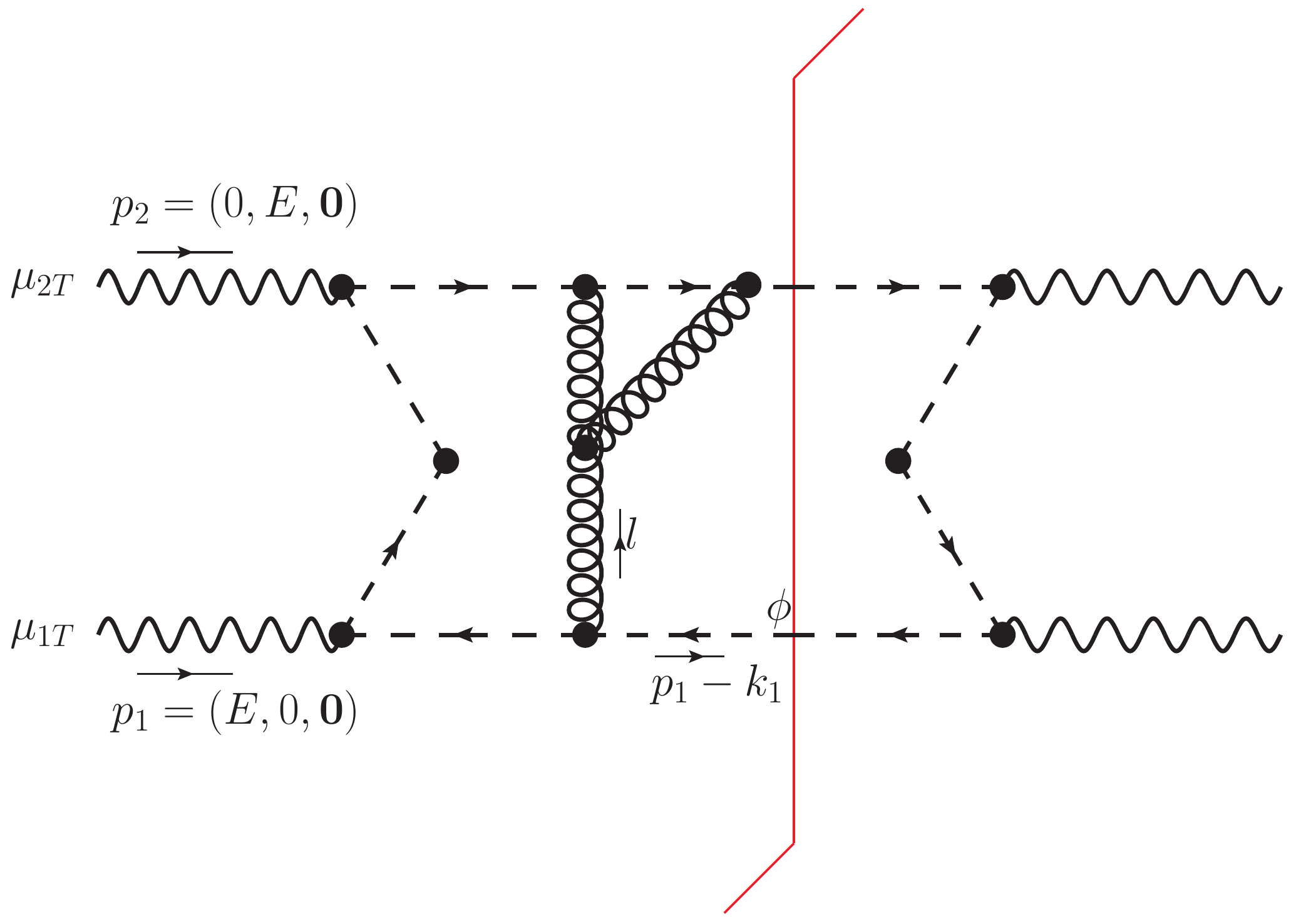} \\
(c) & (d) \\[6pt]
\end{tabular}
\caption{Example cut diagrams, each with only one gluon attached to the spectator line on the lower half of the graph, while more than one gluons may attach to the upper spectator line.}
\label{fig:4figures}
\end{figure}

Consider any cut diagram whose lower spectator line has only one gluon attachment, with the gluon being either soft or Glauber-like, such as the diagrams shown in Fig.\@ \ref{fig:4figures}. As is the case for the one-gluon diagram in Fig.\@ \ref{fig:spectator}, the lower collinear sector only depends on the following three momenta, $p_1$, $k_1$, and $l$. It is also easily checked that, again, at leading power, the only numerator factors that depends on $l_T$ (or the other unlabeled transverse loop momenta) are the photon-scalar vertices.

Therefore, exactly the same transformation as in Sec. \ref{sec:one-Glauber}, Eq.\@ \eqref{eq:reflect}, reverses the sign of the cut amplitude, and proves that the contribution is zero after phase space integration for $k_1$ and $k_2$.
\begin{figure}[tb]
\begin{centering}
\includegraphics[scale=0.5]{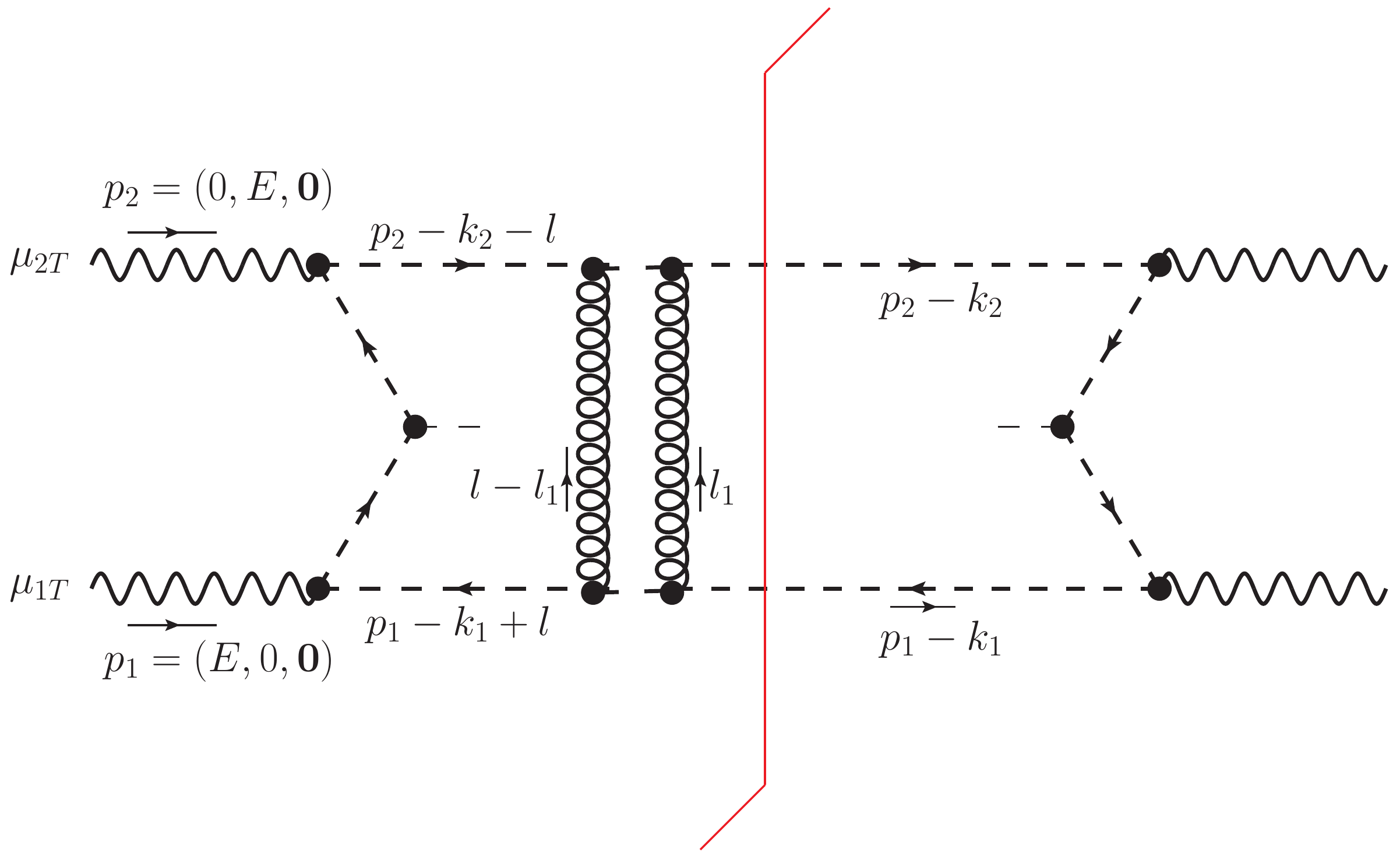}
\par\end{centering}
\caption{Two Glauber gluons exchanged on the same side of the cut, in the model field theory.} 
  \label{fig:spectator-2}
\end{figure}

To prove that the cut diagram in Fig.\@ \ref{fig:spectator-2}, i.e. a 2-loop diagram with a box in interference with the complex conjugate of the LO diagram, vanishes, we need a little more work.
Recall that a ``pinch" in the Glauber region arises when both the active parton and spectator lines depend on the Glauber-like exchanged momentum. Since the active quark line in Fig.\@ \ref{fig:spectator-2} depends on $l$, but not $l_1$, only the overall exchanged momentum $l$ can be pinched in the Glauber region $\sim \left( \lambda^2, \lambda^2, \lambda \right)$, to produce a potentially non-factorizable contribution. Meanwhile, it can be checked by IR power counting that the individual momenta $l_1$ and $l-l_1$ can be both Glauber-like, or both soft $\sim \left( \lambda, \lambda, \lambda \right)$, for the diagram to give a leading-power contribution.\footnote{If $l$ is Glauber-like, $l_1$ and $l-l_1$ can also be both plus-collinear or both minus-collinear for the diagram to contribute at leading power. But it is not necessary to consider this situation, because the sum of diagrams involving a secondary hard vertex is suppressed by Ward identities, as shown by Labastida and Sterman \cite{Labastida:1984gy}.}

As is the case for diagram Fig.\@ \ref{fig:spectator}, the only leading-power dependence of numerator factors on $l_T$ is from the photon-scalar vertices. Here these vertices only depend on $l_T$ but not $l_{1T}$. So the leading-power cut-amplitude is, omitting the color factor $C_F^2$ and other constant factors,
\begin{align}
&\quad \epsilon_{\mu_{1T} \, \nu_{1T}} \epsilon_{\mu_{2T} \, \nu_{2T}} \mathcal M_2^{ \mu_{1T} \, \mu_{2T} } \left( \mathcal M_{\rm LO}^{ \nu_{1T} \, \nu_{2T} } \right)^* \nonumber \\
& \propto \int d^4 l\, \epsilon \left( k_1^T, l^T \right) \epsilon \left(k_2^T, l^T \right) D \left( k_1 - l \right) D \left( k_2 + l \right) D \left( p_1 - k_1 +l \right) D \left( p_2 - k_2 -l \right) \nonumber \\
&\quad  \int d^4 l_1 \, D \left( l_1 \right) D \left( l-l_1 \right) D \left( p_1 - k_1 +l_1 \right) D \left( p_2 - k_2 -l_1 \right), \nonumber \\
\label{eq:LO-2}
\end{align}
where we have collected all the $l_1$-dependent terms in the third line.

We briefly comment on the ``Glauber-II" region discussed in \cite{Stewart:2009yx}, which has momentum scaling $\left( \lambda^4, \lambda^4, \lambda^2 \right)$ instead of the usual Glauber scaling $\left( \lambda^2, \lambda^2, \lambda \right)$. The factor $\epsilon \left( k_1^T, l^T \right) \epsilon \left(k_2^T, l^T \right)$ in Eq.\@ \eqref{eq:LO-2} and Eq.\@ \eqref{eq:LO-1}, absent in unpolarized scattering, gives a suppression when $l_T$ is smaller than the usual Glauber transverse momentum. Therefore, even if the ``Glauber-II" region is relevant for leading-power unpolarized scattering, it is not relevant here. Also there will be no singularity from $l_T \to 0$, which is important because the next step is analyzing the cut diagram at fixed $l_T$, assuming that the subsequent integration over $l_T$ causes no complication.

With $l$ being Glauber-like and $l_1$ being either Glauber-like or soft, the only propagators that have leading-power dependence on $l^-$ are the two lines immediately connected to the lower incoming hadron $p_1$, and the only propagators that have leading-power dependence on $l^+$ are the two lines immediately connected to the upper incoming hadron $p_2$.
So at any fixed $l_T$, we can perform the $l^+$ and $l^-$ integrals by contour integration \cite{Bodwin:1984hc, Collins:2011zzd}, picking up the poles by cutting the lines $p_2-k_2-l$ and $p_1-k_1+l$ in Fig.\@ \ref{fig:spectator-2}, producing an imaginary contribution multiplied by a one-loop box (sub-) diagram initiated by the on-shell lines $p_2-k_2-l$ and $p_1-k_1+l$. To obtain a real contribution that is not canceled by the complex conjugate cut diagram, we need another imaginary contribution from applying cutkosky's rules to the one-loop box diagram, with $l_1$ in the Glauber region.\footnote{As shown in \cite{Donoghue:2014mpa}, applying cutkosky's rules to the direct box diagram alone yields the same imaginary part as applying the method of regions to the sum of the direct box and the cross box diagram.} The approximations are
\begin{align}
D(l_1) &\approx \frac i { -\left| l_{1T} \right|^2 + i0}, \\
D(l-1_1) &\approx \frac i {-\left| l_T-l_{1T} \right|^2 + i0}, \\
D(p_1 - k_1 + l_1) &\approx \frac i {(1-x)Q \left( \frac{ \left| k_{1T} \right|^2 } {(1-x)Q} + l_1^- \right) - \left| k_{1T} - l_{1T} \right|^2 + i0}, \\
D(p_2 - k_2 - l_1) &\approx \frac i {(1-x)Q \left( \frac{ \left| k_{2T} \right|^2 } {(1-x)Q} - l_1^+ \right) - \left| k_{2T} - l_{1T} \right|^2 + i0}
\end{align}
Following Cutkosky's rules, we replace $D(p_1 - k_1 + l_1)$ and $D(p_2 - k_2 - l_1)$ by corresponding delta functions, and integrate over $l_1^-$ and $l_1^+$, producing the constant factor $(2\pi i)^2 / \left[ (1-x)Q \right]^2$. Now the last line of Eq.\ \eqref{eq:LO-2} becomes just the two-dimensional $\int d^2 l_{1T}$ integral over $D(l_1)$ and $D(l-l_1)$. At this point, we can re-use the strategy for showing the vanishing of the interference between Fig.\@ \ref{fig:spectator} and the LO graph, and replace $k_1^T$ in the first line of Eq.\ \eqref{eq:LO-2} by its mirror image in the line through the origin in the direction of $\pm l^T$, as in Eq.\ \eqref{eq:reflect}. The integrand reverses sign, and since this transformation does not affect the doubly differential beam thrust distribution, the final contribution from the cut diagram Fig.\@ \ref{fig:spectator-2} is zero.
\section{Non-zero two-Glauber diagram}
\label{sec:nonzero}
\begin{figure}[tb]
\begin{centering}
\includegraphics[scale=0.5]{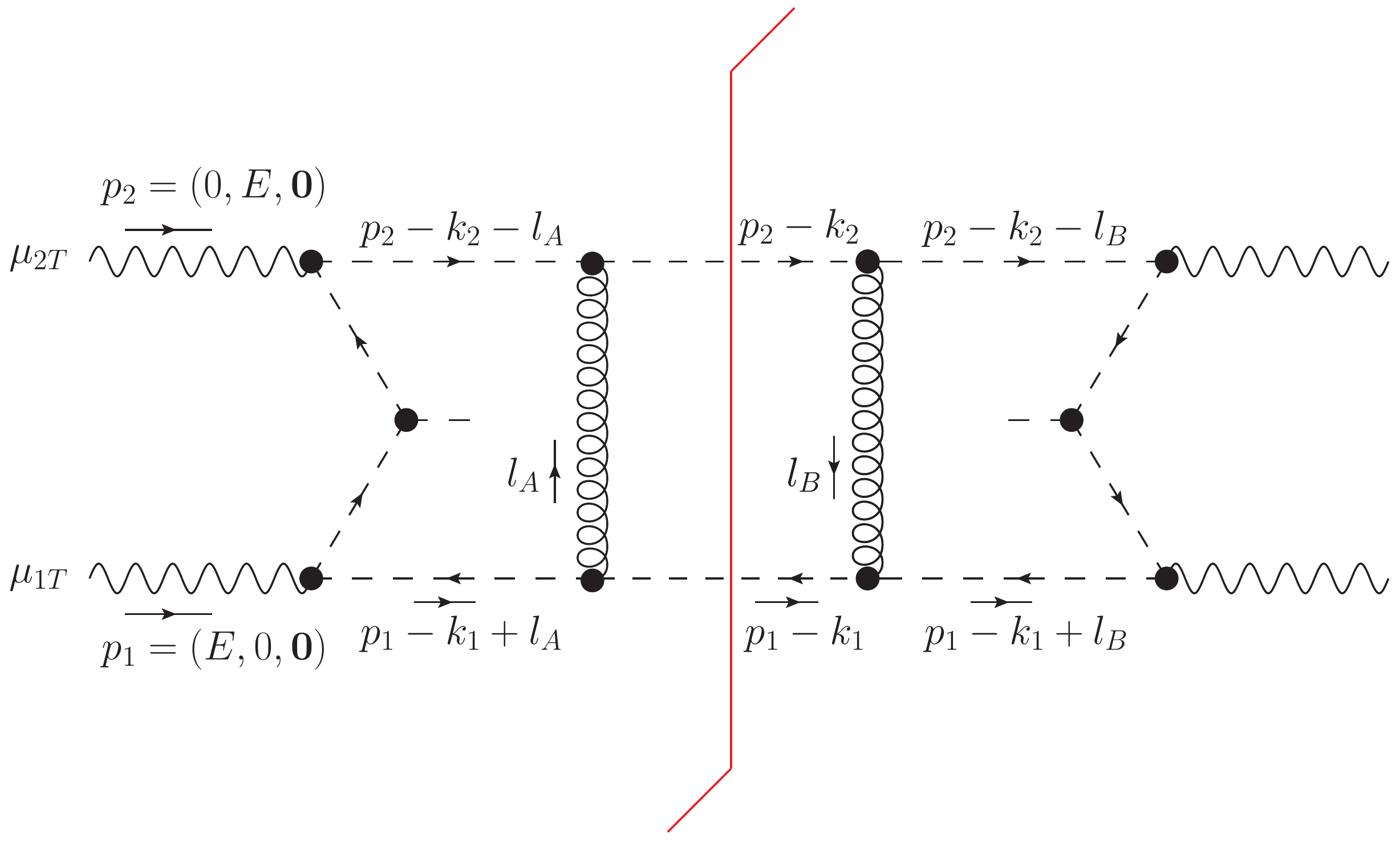}
\par\end{centering}
\caption{The cut diagram with one Glauber gluon exchanged on either side of the cut, i.e.\@ the square of Fig.\@ \ref{fig:spectator}.} 
  \label{fig:nonzero}
\end{figure}

We have excluded all cut diagrams whose lower (or upper) spectator line is attached by only one soft gluon (normal soft gluon or Glauber gluon) in Section \ref{sec:2loopzero}. We have also excluded the diagram involving two Glauber gluons on the same side of the cut, including the diagram with a box, Fig.\@ \ref{fig:spectator-2}, and an unshown diagram with a cross box. This leaves us, at the order $\alpha_s^2$ relative to LO, the only possible contribution from the cut diagram in Fig.\@ \ref{fig:nonzero}, showing the interference of Fig.\@ \ref{fig:spectator} with its own complex conjugate. Out of the two soft gluons exchanged on the two sides of the cut, one has to be a Glauber gluon to produce factorization-violating effects, and then the other one also has to be a Glauber gluon to produce a real contribution that is not canceled by the complex conjugate cut diagram \cite{Bodwin:1984hc, Gaunt:2014ska}.

\subsection{Reducing to 2D integrals by contour integration}
We show that the cut diagram Fig.\@ \ref{fig:nonzero}, from squaring the amplitude in Fig.\@ \ref{fig:spectator}, gives a non-zero contribution to the double longitudinal spin asymmetry.

As above, we fix the heavy particle $\Phi$ to have zero rapidity, so that in Fig.\@ \ref{fig:nonzero}, $x_1=x_2=x,\ k_1^+ = k_2^- = x Q$, and the amplitude Eq.\@ \eqref{eq:M1} can be re-written as
\begin{align}
i \mathcal M_1^{ \mu_{1T} \, \mu_{2T} } 
& = -4i e^2 g_\Phi \, g_s^2 (1-x)^2 Q^2  \int \frac{dl^+ dl^- d^2 l_T}{(2\pi)^4} \, \left( k_1^{\mu_{1T}} - l^{\mu_{1T}} \right) \, \left( k_2^{\mu_{2T}} + l^{\mu_{2T}} \right) \nonumber \\
&\quad D \left( l \right) D \left( p_1 - k_1 +l \right) D \left( p_2 - k_2 -l \right)
D \left( k_1 - l \right) D \left( k_2 + l \right),
\label{eq:M1-again}
\end{align}
In this expression, with $l$ lying in the Glauber region, the relevant leading-power approximations are
\begin{align}
D(l) &\approx \frac i {-l_T^2 + i0}, \\
D(p_1 - k_1 + l) &\approx \frac i {(1-x)Q \left( \frac{ \left| k_{1T} \right|^2 } {(1-x)Q} + l^- \right) - \left| k_{1T} - l_{T} \right|^2 + i0},\\
D(k_1 - l) &\approx \frac i {xQ \left( -\frac{ \left| k_{1T} \right|^2 } {(1-x)Q} - l^- \right) - \left| k_{1T} - l_{T} \right|^2 + i0},\\
D(p_2 - k_2 - l) &\approx \frac i {(1-x)Q \left( \frac{ \left| k_{2T} \right|^2 } {(1-x)Q} - l^+ \right) - \left| k_{2T} - l_{T} \right|^2 + i0},\\
D(k_2 - l) &\approx \frac i {xQ \left( -\frac{ \left| k_{2T} \right|^2 } {(1-x)Q} - l^+ \right) - \left| k_{2T} - l_{T} \right|^2 + i0},
\end{align}
We first integrate over $l^+$ and $l^-$ using contour integration, picking up the poles from the vanishing of $D(p_1 - k_1 + l)$ and $D(p_2 - k_2 - l)$. We are left with the $l_T$ integral,
\begin{align}
i \mathcal M_1^{ \mu_{1T} \, \mu_{2T} } 
& = 4 e^2 g_\Phi \, g_s^2 (1-x)^2 \int \frac{d^2 l_T}{(2\pi)^2} \, \left( k_1^{\mu_{1T}} - l^{\mu_{1T}} \right) \, \left( k_2^{\mu_{2T}} + l^{\mu_{2T}} \right) \nonumber \\
&\quad \frac 1 { \left| l_T \right|^2} \, \frac 1 { \left| k_{1T} - l_T \right|^2} \, \frac 1 { \left| k_{2T} + l_T \right|^2},
\label{eq:M1-lT}
\end{align}
So the asymmetry from the cut diagram Fig.\@ \ref{fig:nonzero}, using the spin sum formula Eq. \eqref{eq:absoluteAsym}, is
\begin{align}
\left| \mathcal M_1 \right|^2_{\rm asym} &= 2 C_F^2 C_A \int \frac {d^2 l_A}{ (2\pi)^2 } \int \frac {d^2 l_B}{ (2\pi)^2 } \frac 1 4 \epsilon_{\mu_{1T} \, \nu_{1T}} \epsilon_{\mu_{2T} \, \nu_{2T}} \mathcal M_1^{ \mu_{1T} \, \mu_{2T} } \left( \mathcal M_1^{ \mu_{1T} \, \mu_{2T} } \right)^* \nonumber \\
&= 2 C_F^2 C_A \cdot 4 e^4 g_\Phi^2 \, g_s^4 (1-x)^4 \int \frac {d^2 l_A}{ (2\pi)^2 } \int \frac {d^2 l_B}{ (2\pi)^2 } \mathcal I_{\rm asym} \left( k_{1T}, k_{2T}, l_A, l_B \right),
\label{eq:M2asym}
\end{align}
where, using the notation in Eq.\@ \eqref{eq:epsilon},
\begin{align}
\mathcal I_{\rm asym} \left( k_{1T}, k_{2T}, l_A, l_B \right) &=
\epsilon \left( k_1-l_A, k_1-l_B \right) \epsilon \left( k_2+l_A, k_2+l_B \right) \nonumber \\
&\quad \times \frac 1 {\left| l_A \right|^2} \frac 1 {\left| k_{1T}-l_A \right|^2} \frac 1 {\left| k_{2T}+l_A \right|^2} \frac 1 {\left| l_B \right|^2} \frac 1 {\left| k_{1T}-l_B \right|^2} \frac 1 {\left| k_{2T}+l_B \right|^2}.
\label{eq:Asym}
\end{align}
In Eq.\@ \eqref{eq:M2asym}, the previously ignored color factor $C_F^2 C_A$ is shown, and an overall factor of 2 is present to account for the possibility of reversing the complex scalar arrow (i.e.\@ swapping scalar and anti-scalar) in Fig.\@ \ref{fig:spectator}.
Using Eq. \eqref{eq:M2asym} as $| \mathcal M |^2$ in Eq. \eqref{eq:phaseSpace1}, we obtain
\begin{equation}
\frac {d^3 \sigma_{\rm asym}} {d \tau_R d \tau_L dy} \bigg|_{y=0, \tau_R = \tau_L = \tau_B/2} = \frac {M^2} {16\pi E^2} 2 C_F^2 C_A \cdot 4e^4 g\Phi^2 \, g_s^4 (1-x)^4 I_{\rm asym},
\label{eq:NNLOasym}
\end{equation}
where we defined $I_{\rm asym}$ as the loop and phase space integral over the integrand, Eq. \eqref{eq:Asym},
\begin{align}
I_{\rm asym} &= \int \frac {d^2 l_A} { (2\pi)^2 } \int \frac {d^2 l_B} { (2\pi)^2 } \int \frac {d^2 k_{1T}} { (2\pi)^2 } \int \frac {d^2 k_{2T}} { (2\pi)^2 } \nonumber \\
&\quad \delta \left( \left| k_{1T} \right|^2 - \left| k_{0T} \right|^2 \right) \delta \left( \left| k_{2T} \right|^2 - \left| k_{0T} \right|^2 \right) \mathcal I_{\rm asym} \left( k_{1T}, k_{2T}, l_A, l_B \right).
\label{eq:integral}
\end{align}
\subsection{Cancellation of IR divergences and numerical evaluation}
Consider the $l_A$ and $l_B$ integrals in Eq.\@ \eqref{eq:integral}. Possible IR divergences may arise from the vanishing of any of the denominators in Eq.\@ \eqref{eq:Asym}, if the singularities are not integrable. The singularities at $l_{A,B} = k_{1T}$ and $l_{A,B} = -k_{2T}$ are integrable, because the numerator factors on the 2nd line of Eq.\@ \eqref{eq:Asym} vanish at these points. However, there seems to be logarithmic singularities when $l_A$ or $l_B$ approaches zero. We show that this is not the case, because there is a linear suppression when $l_A$ or $l_B$, becomes small, turning the singularities into integrable ones.
The subsequent argument will be similar to Section \ref{sec:one-Glauber}, where we used the reflection of $k_1$
to show that the one-Glauber diagram vanishes.
This suppression also implies that there is no leading power contribution from any regions with much smaller transverse momenta, such as the ultra-soft region and the ``Glauber-II" region discussed in \cite{Stewart:2009yx}, or from the overlap between these regions and the regular Glauber region.

We use $R_{p} \circ q$, defined in Eq.\@ \eqref{eq:reflect}, to denote the reflection of the two-vector $q$ with respect to to the line through the origin in the direction of the two-vector $p$. We then re-write Eq.\@ \eqref{eq:integral} as
\begin{align}
I_{\rm asym} &= \frac 1 2 \int \frac {d^2 k_{1T}}{ (2\pi)^2 } \int \frac {d^2 k_{2T}}{ (2\pi)^2 }\, \delta \left( \left| k_{1T} \right|^2 - \left| k_{0T} \right|^2 \right) \delta \left( \left| k_{2T} \right|^2 - \left| k_{0T} \right|^2 \right) \nonumber \\
&\quad \times \int \frac {d^2 l_A}{ (2\pi)^2 } \int \frac {d^2 l_B}{ (2\pi)^2 } \big[ \mathcal I_{\rm asym} \left( k_{1T}, k_{2T}, l_A, l_B \right) + \mathcal I_{\rm asym} \left( R_{l_A} \circ k_{1T}, k_{2T}, l_A, l_B \right) \big].
\label{eq:integral1}
\end{align}
In this form, the last line can be readily checked to vanish when $l_B$=0, and linearly suppressed when $l_B$ is small. The Jacobian factor from the reflection is 1, so we simply need to put a factor of $1/2$ at the start of Eq.\@ \eqref{eq:integral1}. We go one step further by reflecting $k_{2T}$ with respect to the line through $l_B$, recasting Eq.\@ \eqref{eq:integral} into
\begin{align}
I_{\rm asym} &= \frac 1 4 \int \frac {d^2 k_{1T}}{ (2\pi)^2 } \int \frac {d^2 k_{2T}}{ (2\pi)^2 }\, \delta \left( \left| k_{1T} \right|^2 - \left| k_{0T} \right|^2 \right) \delta \left( \left| k_{2T} \right|^2 - \left| k_{0T} \right|^2 \right) \nonumber \\
&\quad \times \int \frac {d^2 l_A}{ (2\pi)^2 } \int \frac {d^2 l_B}{ (2\pi)^2 } \big[ \mathcal I_{\rm asym} \left( k_{1T}, k_{2T}, l_A, l_B \right) + \mathcal I_{\rm asym} \left( R_{l_A} \circ k_{1T}, k_{2T}, l_A, l_B \right) \nonumber \\
&\quad \mathcal I_{\rm asym} \left( k_{1T}, R_{l_B} \circ k_{2T}, l_A, l_B \right) + \mathcal I_{\rm asym} \left( R_{l_A} \circ k_{1T},  R_{l_B} \circ k_{2T}, l_A, l_B \right) \big],
\label{eq:integral2}
\end{align}
where the sum inside the square bracket receive a linear suppression when either $l_A$ or $l_B$ become small, and a quadratic suppression when both $l_A$ and $l_B$ are made small simultaneously. This makes both the points $l_A,\ l_B=0$ integrable singularities despite the quadratic denominators in Eq.\@ \eqref{eq:Asym}. So the expression Eq.\@ \eqref{eq:integral2} is IR finite. UV finiteness is also clear by power counting. With both IR and UV divergences absent, Eq.\@ \eqref{eq:integral2} can be evaluated by straightforward Monte Carlo integration without regularization or subtraction. Using the Vegas algorithm implemented by the CUBA library \cite{Hahn:2005pf}, with $4.2$ million points sampled, we obtain
\begin{equation}
I_{\rm asym} =(1.58 \pm 0.02) \frac {1} { (4\pi)^4 \left| k_{0T} \right|^4 }
\label{eq:asym-final}
\end{equation}
Dividing the absolute asymmetry given by Eqs.\@ \eqref{eq:NNLOasym} and \eqref{eq:asym-final} by the LO unpolarized differential cross section given by Eqs.\@ \eqref{eq:LOdistribution} and \eqref{eq:integralLO}, we obtain the relative spin asymmetry
\begin{align}
\left( \frac {d^3 \sigma_{\rm asym}} {d \tau_R d \tau_L dy} / \frac {d^3 \sigma_{\rm LO}} {d \tau_R d \tau_L dy} \right)
\bigg|_{y=0, \tau_R = \tau_L = \tau_B/2}
&= C_F^2 g_s^4 \cdot I_{\rm asym} / I_{\rm LO} \nonumber \\
&=(1.58 \pm 0.02) C_F^2 \alpha_s^2.
\label{eq:relative-final}
\end{align}
We have shown that the asymmetry is non-zero, proving that Glauber gluons break factorization for the double spin asymmetry in the doubly differential beam thrust distribution.

\section{Discussion}
\label{sec:dis}
We have performed a calculation of factorization-violating effects in the beam thrust distribution from Drell-Yan-like scattering in a simple model field theory. Any factorization in the limit of small beam thrust (corresponding to a stringent jet veto), standard or generalized, would predict a vanishing double longitudinal spin asymmetry, due to the scalar nature of the active quarks in this parity-conserving model.
The non-zero result found in our calculation is in contradiction to any generalized factorization that separates beam-thrust dependence into universal functions. The non-factoring contribution, Eq.\ \eqref{eq:relative-final}, is in fact infrared safe, which shows that collinear factorization is respected to this order, but with factorization scale $\mu_F = \mathcal O ( \sqrt {\tau_B s} )$. Logarithms of beam thrust are thus contained in the hard-scattering function of collinear factorization, and standard resummation methods do not reply.

By looking at double spin asymmetry in a theory with scalar quarks, the calculation is simplified enormously. For example, the diagrams in Fig.\ \ref{fig:4figures} and \ref{fig:spectator-2} all vanish, and the only non-zero diagram Fig.\ \ref{fig:nonzero} is IR-finite in the Glauber region without regularization. Neither of these simplifications hold for unpolarized scattering in real QCD. Since the factorization theorem in \cite{Stewart:2009yx} for the beam thrust distribtuion is proposed in a manner that applies to both unpolarized and polarized scattering in any unbroken gauge theory, our result is a counter-example which shows the aforementioned factorization theorem cannot hold true in every situation. But strictly speaking, we do not exclude the small possiblity that the factorization theorem survives in unpolarized scattering; a dedicated, more complicated calculation for unpolarized scattering would be needed to conclusively settle the question.

The breakdown of generalized factorization would eventually lead to corrections to the existing predictions of jet veto resummation calculations \cite{Banfi:2012yh, Banfi:2012jm, Stewart:2013faa, Tackmann:2012bt, Becher:2012qa, Becher:2013xia}. The question remains, \emph{``At which logarithmic order do such corrections start?"} The lowest-order factorization-violating diagram in this study involves two spectator lines and two virtual Glauber gluons, producing a non-zero result. This would be of order $\alpha_s^4$ if we were studying massless parton scattering instead of photon-photon scattering. This result contains no large logarithms for two reasons. First, the intrinsic virtuality of collinear particles, anti-collinear particles, and Glauber gluons are all of the order $Q^2 \tau_B$ \cite{Stewart:2009yx}. Second, the order of the diagram is too low to acquire Regge-type rapidity logarithms, which will show up at higher orders in ladder-type diagrams. But we still need to multiply the result by the hard function (which always factorizes, though collinear and soft functions can be entangled by Glauber gluons), with double logarithms $\sim \alpha_s^n \ln^{2n} \tau_B$ due to running from the scale $Q^2$ to $Q^2 \tau_B$. We end up with $\alpha_s^{n+4} \ln^{2n} \tau_B$. In unpolarized scattering, there can also be a non-zero contribution when two soft / Glauber gluons are exchanged on the same side of the cut, which potentially gives one power of a Regge-type logarithm, resulting in $\alpha_s^{n+4} \ln^{2n+1} \tau_B$. This suggests a breakdown of naive jet veto resummation at no later than ${\rm N}^4$LL.

A corollary of the study is that a proper description of the Drell-Yan process with stringent jet vetoes must include entanglement between the two collinear sectors. It should be noted that our study only demonstrates the inevitability of entangling the two collinear sectors, while soft and ultra-soft gluons may still be allowed to factorize in some manner.

Discrete symmetries play an important role in our approach. If factorization holds, the two collinear sectors are decoupled and have separate parity invariance, resulting in $Z_2 \times Z_2$ symmetry, which is richer than the $Z_2$ global parity symmetry of QCD. Violation of factorization is revealed by the violation of extra discrete symmetries resulting from factorization. To construct a more realistic example that could be tested at colliders with unpolarized beams, one could exploit charge conjugation invariance: for example, the proton and the anti-proton have the same gluon beam function. If a future calculation shows that the Higgs production cross section (in the gluon fusion channel only) under a stringent jet veto in $pp$ collisions is different, at leading power in $\mathcal O (p_T^{\rm veto} / M_{\rm higgs} )$, from the same quantity in $p\bar p$ collisions, it would be another manifestation of the violation of factorization.
\section{Acknowledgment}
The author would like to thank George Sterman for enlightening discussions. The Jaxodraw program \cite{Binosi:2003yf} is used to draw Feynman diagrams. The work is supported by the NSF grant PHY-1316617.

\appendix
\section{Definition of the beam function using QCD fields}
\label{sec:beam}
For completeness, we give a definition of the beam function using QCD fields. This appendix is essentially a review and does not contain original work, because the definition agrees with the SCET definition \cite{Stewart:2009yx} at least at low orders \cite{Lee:2006nr,Idilbi:2007ff,Idilbi:2007yi,Feige:2015rea}.
The \emph{unsubtracted} momentum-space beam function for a scalar parton $\phi$ with Bjorken variable $x_1$ and virtuality (ignoring transverse momentum components) $\omega_1 = x_1 P^+ M \tau_{R}$, for an incoming hadron $| H_1 \rangle$ with a large plus momentum component $P^+$,
\begin{align}
B_1^{\rm unsubtracted} \left( \omega_1, x_1, \mu \right) &= \frac{x P^+}{2} \int \frac{dw^-} {2\pi} \frac{dw^+} {2\pi} e^{-i \left( x_1 P^+ w^- + M \tau_{R}\, w^+  \right)/2 } \langle H_1 | \phi^\dagger \left( w^+, w^-, \bm 0_T \right) \nonumber \\
&\quad \mathcal P \exp \left[ \int_{0}^{w^-} \frac{dy^-}{2} i g A^+ \left(0, y^-, \bm 0_T \right) \right] \phi (0) | H_1 \rangle,
\end{align}
where $\mu$ is the UV renormalization scale at which the matrix element is defined.
This is directly analogous to Eq.\ (50) in \cite{Stewart:2009yx} (with ``$+$'' and ``$-$'' exchanged), except that the latter reference used the SCET collinear field with zero-bin subtraction.

The eikonal beam function is defined by replacing the incoming hadron and the interpolating field by a Wilson line in the ``$+$" direction,
\begin{align}
B_1^{\rm eikonal} \left( \omega_1, x_1, \mu \right) &= \frac{x P^+}{2} \int \frac{dw^-} {2\pi} \frac{dw^+} {2\pi} e^{-i \left( x_1 P^+ w^- + M \tau_{R}\, w^+  \right)/2 } \nonumber \\
& \quad \langle 0 |\, W_1^\dagger \left( w^+, w^-, \bm 0_T \right) \mathcal P \exp \left[ \int_{0}^{w^-} \frac{dy^-}{2} i g A^+ \left(0, y^-, \bm 0_T \right) \right] W_1(0)\, | 0 \rangle,
\end{align}
where we defined
\begin{equation}
W_1 \left( w^+, w^-, \bm w_T \right) = \mathcal P \exp \left[ \int_{-\infty}^{0} \frac{dy^+}{2} i g A^- \left( w^+ + y^+, w^-, \bm w_T \right) \right].
\end{equation}

Finally, we divide the Laplace transform of $B_1^{\rm unsubtracted}$ with respect to $\omega$, $\tilde B_1^{\rm unsubtracted}$, by the Laplace transform of $B_1^{\rm eikonal}$, $\tilde B_1^{\rm eikonal}$, to obtain the gauge-invariant moment-space beam function $\tilde B_1 \left(\tilde \omega, x_1, \mu \right)$. The inverse Laplace transform of this result is the momentum-space beam function $B_1 \left( \omega, x_1, \mu \right)$.


\bibliographystyle{JHEP}
\bibliography{glauber}
\end{document}